\documentclass[twocolumn,floatfix,noeprint,superscriptaddress]{revtex4-1}

\usepackage{dsfont} 
\usepackage{bm} 

\usepackage{bbold} 
\usepackage{xcolor}
\usepackage{times}
\usepackage{float}
\usepackage{physics}
\usepackage{graphicx,braket}
\usepackage{hyperref}
\hypersetup{colorlinks=true,linkcolor=blue,citecolor=blue}
\usepackage{amsmath,amssymb,amsfonts}
\usepackage{wrapfig}

\usepackage{cleveref} 

\usepackage{siunitx}

\usepackage{cleveref}
\crefname{equation}{Eqs.}{Eqs.}
\Crefname{equation}{Equation}{Equations}
\crefrangelabelformat{equation}{(#3#1#4--#5#2#6)}
\crefmultiformat{equation}{Eqs. (#2#1#3}{, #2#1#3)}{#2#1#3}{#2#1#3}
\Crefmultiformat{equation}{Equations (#2#1#3}{, #2#1#3)}{#2#1#3}{#2#1#3}
\usepackage{ amssymb }
\usepackage{lipsum}
\usepackage{kantlipsum}

\AtBeginDocument{%
    \newwrite\bibnotes
    \def\bibnotesext{Notes.bib}
    \immediate\openout\bibnotes=\jobname\bibnotesext
    \immediate\write\bibnotes{@CONTROL{REVTEX41Control,eprint=""}}
    \immediate\write\bibnotes{@CONTROL{%
    apsrev41Control,author="08",editor="1",pages="1",title="0",year="1"}}
     \if@filesw
     \immediate\write\@auxout{\string\citation{apsrev41Control}}%
    \fi
}%

\begin{document}

\allowdisplaybreaks 

\flushbottom
\title{Two-photon resonance fluorescence of two interacting non-identical quantum emitters}
\author{Alejandro Vivas-Via{\~n}a}
\affiliation{Departamento de Física Teórica de la Materia Condensada and Condensed
Matter Physics Center (IFIMAC), Universidad Autónoma de Madrid, Madrid,
Spain}
\author{Carlos S\'anchez Mu\~noz }
\email[]{carlossmwolff@gmail.com}
\affiliation{Departamento de Física Teórica de la Materia Condensada and Condensed
Matter Physics Center (IFIMAC), Universidad Autónoma de Madrid, Madrid,
Spain}

\newcommand{\down}{\op{g}{e}}
\newcommand{\up}{\op{e}{g}}
\newcommand{\downd}{\op{+}{-}} 
\newcommand{\upd}{\op{+}{-}}
\newcommand{\app}{a^\dagger}
\newcommand{\ssp}{\sigma^\dagger}
\newcommand*{\Resize}[2]{\resizebox{#1}{!}{$#2$}}%

\begin{abstract}
We study a system of two interacting, non-indentical quantum emitters driven by a coherent field. We focus on the particular condition of two-photon resonance and obtain analytical expressions for the stationary density matrix of the system and observables of the fluorescent emission. Importantly, our expressions are valid for the general situation of non-identical emitters with different transition energies. Our results allow us to determine the regime of parameters in which coherent two-photon excitation, enabled by the coherent coupling between emitters, is dominant over competing, first-order processes. Using the formalism of quantum parameter estimation, we show that the features imprinted by the two-photon dynamics into the spectrum of resonance fluorescence are particularly sensitive to changes in the distance between emitters, making two-photon excitation the optimal driving regime for the estimation of inter-emitter distance. This can be exploited for applications such as superresolution imaging of point-like sources.
\end{abstract}
\date{\today} \maketitle

\section{Introduction}

The cooperative phenomena emerging from the interaction between multiple quantum emitters and a common electromagnetic mode is one of  the central topics in quantum optics~\cite{Dicke1954,Ficek2005,Garraway2011,Kirton2019}. The minimal implementation of this paradigm---two quantum emitters---already features the most relevant of these collective effects, e.g. superradiant emission and dark states~\cite{Ficek2005,Lehmberg1970}.  
Minimal models of two and three quantum emitters have been studied extensively in the literature~\cite{Lehmberg1970,RiosLeite1980,Richter1982,Ficek1983,Palma1989,Varada1992,Itano1988,Tanas2004,
Beige1998,Lembessis2013,LAKSHMI2014,Ahmed2014,Peng2019a,Darsheshdar2021,Peng2020,Wang2020}, and examples of emergent phenomenology include superradiance~\cite{DeVoe1996,Mlynek2014}, generation of qubit entanglement~\cite{Gonzalez-Tudela2011,Tanas2004,Alharbi2010} and spin and light squeezing~\cite{Ficek1984,Ficek1994,Haakh2015}, non-classical photon correlations~\cite{Itano1988,Peng2019a,Darsheshdar2021,Peng2020}, emission of entangled photons~\cite{Wang2020}, and potential for molecule localization with nanometer resolution~\cite{Hettich2002,Zhang2016a}.
The insights provided by these minimal theoretical models apply to a large variety of physical systems, including coupled quantum dots~\cite{Gerardot2005,Reitzenstein2006,Laucht2010,Patel2010}, trapped ions~\cite{DeVoe1996,Eschner2001}, Rydberg atoms~\cite{Ates2007,Amthor2010,Pritchard2012}, molecular systems~\cite{Hettich2002,Zhang2016a}, and superconducting qubits~\cite{Lambert2016,Mlynek2014,VanLoo2013}.
Interest on the quantum optical properties of systems of few interacting emitters has been further propelled by the development of photonic nanostructures that mediate and enhance emitter-emitter interactions~\cite{Lodahl2015,Chang2018,Haakh2015,Reitzenstein2006,Laucht2010,Gonzalez-Tudela2011,Mlynek2014,VanLoo2013}.

 Here, we focus on a particularly relevant effect that arises when interacting emitters are driven by a classical  field: the coherent, non-linear excitation of the transition from the ground state to a doubly-excited state via a two-photon resonance, enabled by the emitter-emitter interaction~\cite{Varada1992,Hettich2002}. This mechanism of two-photon excitation, which lies at the heart of important technological applications such as two-photon microscopy~\cite{Zipfel2003}, has attracted great interest for the generation of squeezing~\cite{Ficek1994}, steady state atomic entanglement~\cite{Ficek2002a,Haakh2015}, and emission of entangled photons~\cite{Wang2020}. 
The emergence of an extra peak in the excitation spectrum due to this two-photon resonance has been demonstrated experimentally~\cite{Hettich2002},  used as a method to quantify dipole coupling and, indirectly, estimate the distance between quantum emitters with nanometer resolution.

 Despite the high fundamental and technological relevance of  this mechanism, and the apparent simplicity of the model that describes it, most theoretical studies have focused on the particular case of identical emitters, where one can find analytical expressions for the stationary density matrix of the quantum emitters by direct diagonalization~\cite{Richter1982,Ficek1983,Ficek2002a}. A straightforward analytical solution cannot be obtained in the more complicated case of two non-identical emitters (e.g. with different transition frequencies), where only approximated solutions for steady-state populations, limited to the case of very weak coupling and driving strength have been reported~\cite{Varada1992,Haakh2015}.  The case of non-identical emitters is the common situation in solid-state emitters~\cite{Hettich2002,Patel2010,Haakh2015}, and it is a relevant situation for related models describing, e.g., light-harvesting~\cite{Sanchez2020} and energy-transfer~\cite{Li2015,Zhang2016a}. Given its importance, a full theoretical description of interacting non-identical quantum emitters under coherent driving is desirable. 

In this work, we obtain general analytical expressions for the stationary density matrix of two interacting non-identical quantum emitters under coherent driving at the two-photon resonance. Our expressions are valid for an ample regime of parameters, under the only condition that the energy splitting between one-excitation eigenstates must be the largest energy scale in the system. This allows us to provide closed-form expressions in regimes that could not be described by previous analytical results~\cite{Varada1992}, such as large driving strengths that saturate the two-photon transition, and large coupling between quantum emitters. 
Furthermore, motivated by the experimental work in~\cite{Hettich2002}, we establish the potential of resonance fluorescence measurements to estimate the distance between dipoles below Abe's limit of difraction. Using the framework of quantum parameter estimation~\cite{PARIS2009,Wiseman2009,Dowling2015,PETZ2011,Luis2012,Chao2016,Dominik2018,Liu2020}, we find that the maximum precision is obtained by driving the system at the two-photon transition in the onset of saturation.

This paper is organized as follows. In Section~\ref{sec:1-model}, we describe the model of the system and then introduce two effective models that account for second-order and first-order processes independently. In Section~\ref{sec:2_steady_state}, we apply the results of the previous section to analyze the steady-state observables of the light radiated by the quantum emitters. In Section~\ref{sec:3-spectrum}, we describe the spectrum of two-photon resonance fluorescence. Finally, in Section~\ref{sec:4-parameter-estimation}, we use quantum parameter estimation theory to analyze the potential of spectrum measurements for the estimation of the inter-emitter distance.

\section{Model}
\label{sec:1-model}
\begin{figure*}[t!]
	\begin{center}
		\includegraphics[width=0.99\textwidth]{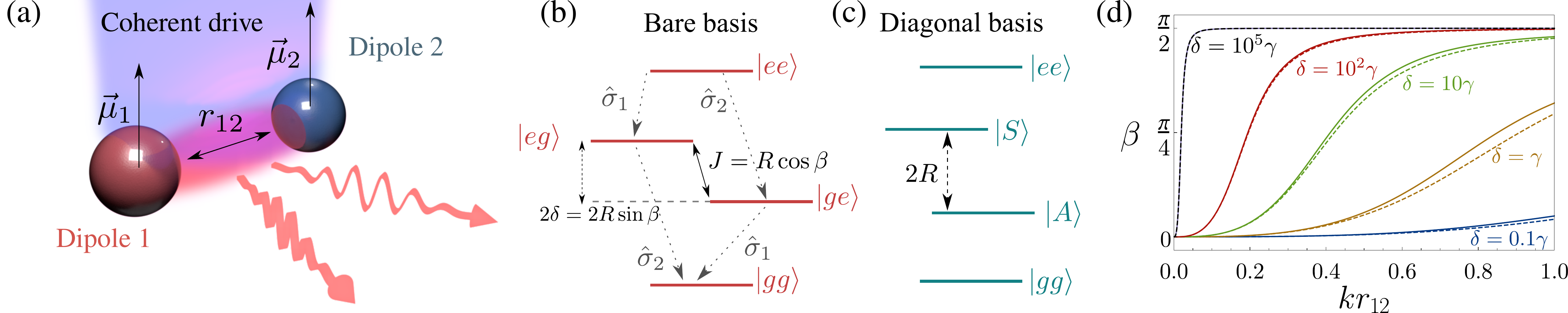}
	\end{center}
\caption{(a) Sketch of the studied system: two interacting non-identical dipoles under coherent driving. (b) Bare basis of the system of quantum emitters. The two states of the one-photon subspace are detuned by an energy $2\delta$, and coupled with a coupling strength $J$. (c) Eigenstates of the Hamiltonian of coupled emitters.(d) Dependence of the 
 mixing angle $\beta = \arctan(\delta/J)$ on the normalized separation between emitters, $k r_{12}$. Dashed lines are obtained via the approximated equation Eq.~\eqref{eq:J-approx}, that assumes $k r_{12}\ll 1$.  From bottom to top the curves correspond to increasing detunings, $\delta/\gamma = {0.1, 1, 10, 100, 10^5}$.}
\label{fig:fig1_setup}
\end{figure*}
\subsection{General model}
Our system is composed of two quantum emitters, each of them described as a two-level system (TLS)  placed at position $\mathbf{r}_i$ ($i\in 1,2$), with natural frequency $\omega_i$ and dipole moment $\bm\mu_i$. The system and energy levels are sketched in Fig.~\ref{fig:fig1_setup}(a-c): each TLS spans a basis $\{|g_i\rangle,|e_i\rangle \}$, where we define the lowering operator $\hat\sigma_i = |g_i\rangle\langle e_i|$. The total basis of the composite system is $\{|gg\rangle, |ge\rangle, |eg\rangle, |ee\rangle\}$, where $|gg\rangle \equiv |g_1\rangle\otimes |g_2\rangle$, and similarly for the other states.  We consider a coherent coupling between both TLS with a coupling rate $J$. In this work, $J$ will be determined by the dipole-dipole interaction between the emitters, although the same model and analysis can be applied in situations in which the interaction between quantum emitters is modified by the mediation of a photonic structure~\cite{Lodahl2015,Chang2018,Haakh2015,Reitzenstein2006,Laucht2010,Gonzalez-Tudela2011,Mlynek2014,VanLoo2013}.
We also include a coherent laser field of frequency $\omega_\mathrm{L}$ driving both emitters with a Rabi frequency $\Omega$. In the rotating frame of the laser, the resulting time-independent Hamiltonian is $\hat H = \hat H_0 + \hat H_\mathrm{d}$, where $\hat H_0$ is the bare Hamiltonian of the interacting quantum emitters
\begin{equation}
\label{eq:H0}
\hat H_0= \left( \Delta - \delta \right) \hat\sigma_1^+ \hat\sigma_1 + 
\left( \Delta +\delta \right) \hat\sigma_2^+ \hat \sigma_2 + J \left(\hat\sigma_1^+ \hat \sigma_2 + \text{h.c.}  \right),  
\end{equation}
and $\hat H_\mathrm d$ is the Hamiltonian of the coherent driving,
\begin{equation}
\hat H_\mathrm{d}= \Omega\left(\hat\sigma_1 + \hat\sigma_2 + \text{h.c.}\right),
\label{eq:Hd}
\end{equation}
having defined $\Delta\equiv(\omega_1 + \omega_2)/2-\omega_L$,  and $\delta\equiv (\omega_2 - \omega_1)/2$, and set $\hbar = 1$. 
Both quantum emitters interact with the electromagnetic field continuum that surrounds them. This field is responsible for mediating the coherent interaction between emitters in Eq.~\eqref{eq:H0}, and also provides a decay mechanism that de-excites the quantum emitters by spontaneous emission to free space. In the reduced Hilbert space of the emitters, this dissipative dynamics  is modeled by a master equation for the density matrix~\cite{Breuer2007,Ficek2005,Gardiner2004},
\begin{equation}
\frac{d\hat\rho}{dt}=-i[\hat H,\hat \rho] +\sum_{i,j=1}^2 \frac{\gamma_{ij}}{2}\mathcal{L}[{\hat\sigma_i,\hat\sigma_j}]
\{\hat\rho\} ,
\label{eq:master_eq}
\end{equation}
where $\mathcal{L}[{\hat O_i,\hat O_j}]
\{\hat\rho\}  \equiv  2 \hat{O}_i \hat\rho \hat{O}_j^+ - \left\{ \hat{O_j}^+ \hat{O_i}, \hat\rho\right\}$, $\gamma_{ii}$ is the local decay rate of spontaneous emission of the $i$-th emitter, and  $\gamma_{12} = \gamma_{21}$ is the dissipative coupling rate between emitters that emerges as a consequence of collective decay. We assume optical transitions, and therefore neglect the incoherent excitation by thermal photons.   The local decay rates depend on each emitter's natural frequency and dipole moment,
\begin{equation}
	\gamma_{ii}\equiv \gamma_i=\frac{\omega_i^3|\bm\mu_i|^2}{3\pi \epsilon_0 \hslash c^3}.
	\label{eq:gamma}
\end{equation}

For the sake of simplicity and without loss of generality, we will assume that both dipole moments equal,  $\bm \mu_1 =\bm \mu_2$. This implies $\gamma_{ii}\approx \gamma$ (assuming $\omega_1,\omega_2 \gg \delta $) and justifies our choice of the same Rabi frequency $\Omega$ for the driving of both emitters,  since, besides having the same dipole moment, their separation  will be considered smaller than the resonant wavelength, $kr_{12}$, and therefore both emitters are driven with the same amplitude, $\Omega(\mathbf r_1) \approx \Omega(\mathbf r_2) \approx \Omega$. Nevertheless, all the results that we obtain can be easily generalized to the case of $\gamma_1 \neq \gamma_2$, $\Omega(\mathbf r_1)\neq \Omega(\mathbf r_2)$. We emphasize that, even if we assume equal dipole moments, we still consider the general case of non-identical emitters that may have different natural  frequencies, i.e. $\delta\neq 0$. The fact that our results apply to this general case is one of the main achievements of this paper.

The coherent and dissipative coupling rates depend on the dipole moments and and also on the relative separation between emitters, ${\mathbf{r}_{12} = \mathbf r_1 -\mathbf r_2}$~\cite{Ficek2005},
\begin{subequations}
\begin{align}
J=
	&\frac{3}{4}\sqrt{\gamma_1 \gamma_2} \left\{ - \left[ 1- \left( \bm{\mu}\cdot \mathbf{r}_{12} \right)^2 \right] \frac{\cos\left( k r_{12} \right)}{kr_{12}} \right. \notag \\
	 &\left. + \left[ 1-3 \left( \bm{\mu}\cdot \mathbf{r}_{12} \right)^2 \right] \left[  \frac{\sin\left( k r_{12} \right)}{(kr_{12})^2} +\frac{\cos\left( k r_{12} \right)}{(kr_{12})^3} \right] \right\},
 \\
\gamma_{12}=
	&\frac{3}{2} \sqrt{\gamma_1 \gamma_2} \left\{ \left[ 1- \left( \bm{\mu}\cdot \mathbf{r}_{12} \right)^2 \right] \frac{\sin \left( k r_{12} \right)}{kr_{12}}  \right. \notag \\
	& \left. + \left[ 1-3 \left( \bm{\mu}\cdot \mathbf{r}_{12} \right)^2 \right] \left[  \frac{\cos\left( k r_{12} \right)}{(kr_{12})^2} - \frac{\sin \left( k r_{12} \right)}{(kr_{12})^3} \right] \right\},
\end{align}
\end{subequations}
where $r_{12}=|\mathbf r_{12}|$, $k = \omega_0/c$, and $\omega_0 = (\omega_1+\omega_2)/2$, having assumed $\omega_0 \gg (\omega_2-\omega_1)$. We are interested in the case of two very close TLS, i.e., $kr_{12}\ll 1$. In this regime, the collective parameters reach their maximal values
\begin{align}
	J \approx
	&\frac{3 \sqrt{\gamma_1 \gamma_2}}{4 (kr_{12})^3}\left[ 1 - 3 (\bm{\mu}\cdot \mathbf{r}_{12})^2 \right],
\label{eq:J-approx}
	 \\[2pt]
	\gamma_{12} \approx
	& \sqrt{\gamma_1 \gamma_2}.
	\label{eq:gamma12-approx}
\end{align}
Henceforth, we adopt ~\cref{eq:J-approx,eq:gamma12-approx} in all our calculations, and also assume that the dipole moments are perpendicular to $\mathbf{r}_{12}$, so ${\bm{\mu} \cdot \mathbf{r}_{12}=0}$. 

The Hamiltonian of the undriven emitters~\eqref{eq:H0} can be easily diagonalized, yielding a new basis $\{|gg\rangle, |A\rangle, |S\rangle, |ee\rangle \}$ where the eigenstates  in single-photon subspace are given by
\begin{subequations}
\begin{align}
|S \rangle&= \frac{1}{\sqrt{2}}\left(\sqrt{1+\sin\beta}|ge\rangle + \sqrt{1-\sin\beta}|eg\rangle \right), \\ 
|A \rangle&= \frac{1}{\sqrt{2}}\left(-\sqrt{1-\sin\beta}|ge\rangle + \sqrt{1+\sin\beta}|eg\rangle \right),
\end{align}
\label{eq:eigenstates}
\end{subequations}
with $\beta$ being  a mixing angle defined as
\begin{equation}
\beta \equiv \arctan(\delta/J).
\label{eq:beta}
\end{equation}
 This diagonal basis is depicted in  Fig.~\ref{fig:fig1_setup}(c). In the case of identical emitters usually discussed in the literature, $\delta=0$ and therefore $\beta = 0$, so that $|S\rangle$ and $|A\rangle$ are, respectively, purely symmetrical and anti-symmetrical superpositions of the states $|ge\rangle$ and $|eg\rangle$ (hence the notation used). In this work, we will consider the more general case where $\beta \in [0,\pi/2]$, which includes the possibility of non-identical emitters, $\delta\neq 0$. The corresponding energies of these two eigenstates are $E_{S/A} = \Delta \pm R$, where we have defined the Rabi frequency of the dipole-dipole coupling as
\begin{equation}
R = \sqrt{J^2 + \delta^2}.
\label{eq:Rabi}
\end{equation}
In this work, we will be particularly interested in the case in which $R$ is the largest energy scale in the system, so that $\hat H_d$ can be treated perturbatively with respect to $R$. This approach is different to the one taken, for instance, in Ref.~\cite{Varada1992}, where $J$ was taken as a perturbative parameter. Our approach will allow  us to derive analytical expressions valid in more regimes, such as the one of high dipole-dipole coupling $J\gg \delta$ that can be relevant for closely spaced emitters. This will also imply that we focus on a regime where the two resonances $\Delta = \pm R$ are well resolved, $R\gg \gamma$, and are therefore visible as two distinct peaks in measurements such as resonance fluorescence excitation spectra~\cite{Hettich2002}.

Since we are interested in using $R$ as our energy reference~\cite{Sanchez2020}, we will reformulate the Hamiltonian parameters in terms of $R$ and $\beta$ as
\begin{equation}
	J=R\cos \beta, \quad \delta= R \sin \beta.
	\label{eq:beta}
\end{equation}
When $J$ is given by the dipole-dipole coupling, as we consider in this work, $\beta$ will depend on both $\delta$ and the distance between emitters, $k  r_{12}$. This makes $\beta$ range from $0$, at short distances, to $\pi/2$, at long distances, as we show in Fig.~\ref{fig:fig1_setup}(d). In this figure, one can see that, as the detuning $\delta$ is decreased, the distance required to reach $\beta=\pi/2$ increases, until one reaches the limiting case $\delta=0$, where $\beta=0$ for any value of the distance $kr_{12}$. The distance $kr_{12}$ has thus as strong impact in the structure of the eigenstates ~\eqref{eq:eigenstates}, which, as we shall see, will affect the quantum optical properties of the emitted light and the response to coherent, two-photon driving. This panel also shows that Eq.~\eqref{eq:J-approx} provides a good approximation for $J$ even for values $k r_{12}\sim 1$.

\subsection{Effective models}
\begin{figure}[b!]
	\begin{center}
 		\includegraphics[width=0.99\columnwidth]{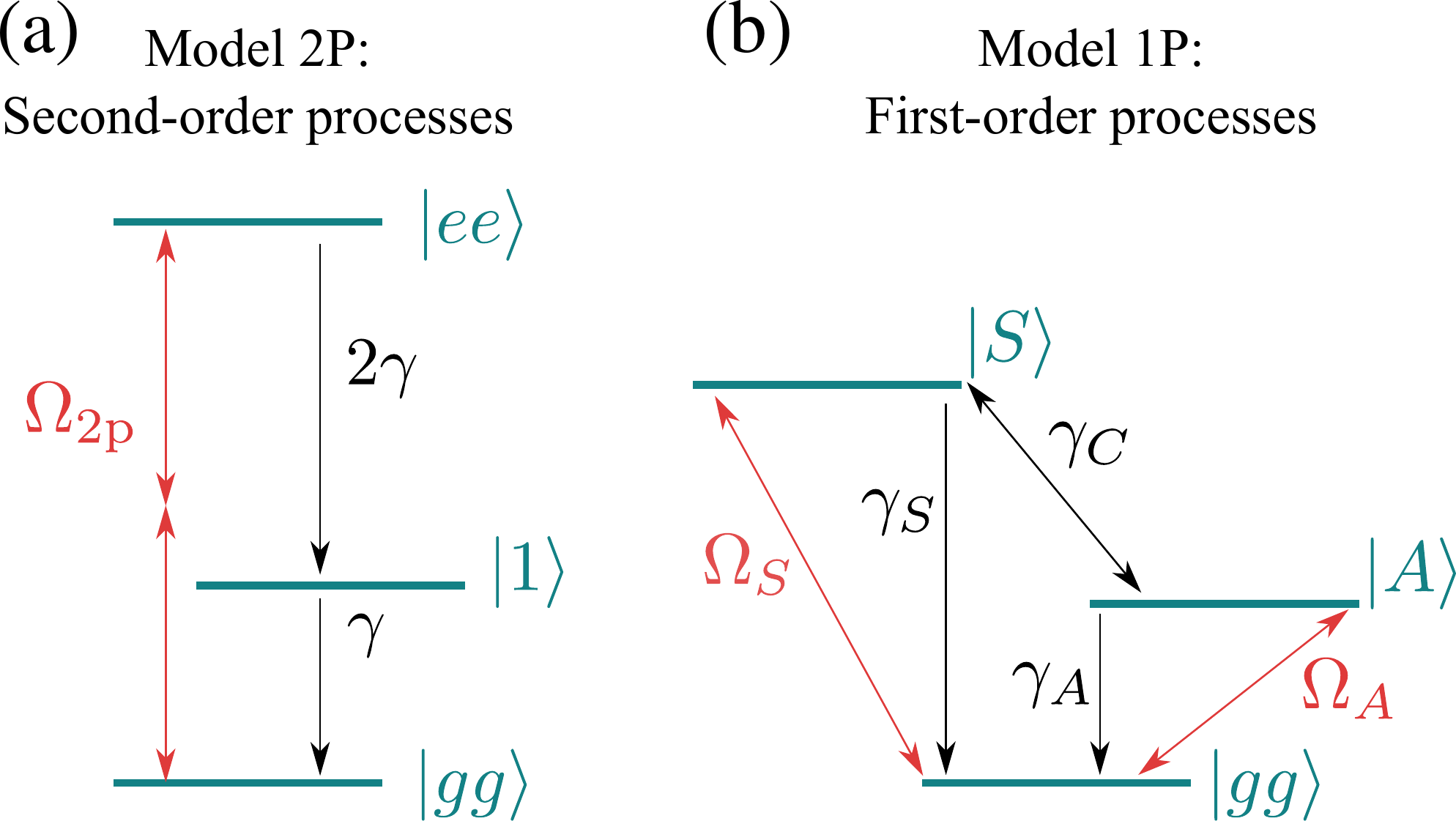}
	\end{center}
\caption{Diagrams of the two effective models used in this work. Red arrows denote coherent coupling between energy levels enabled by the external drive. Black arrows denote channels of irreversible decay. $\gamma_A$, $\gamma_S$ and $\gamma_C$ describe channels of incoherent decay and incoherent coupling in the diagonal basis.}
\label{fig:fig2_effective_models}
\end{figure}

The full master equation in Eq.~\eqref{eq:master_eq} does not yield analytical expressions in general case of $\delta
\neq 0$, which explains why the majority of literature has focused on the case $\delta=0$. In order to obtain analytical expressions for the system density matrix, we will make the assumption that the dynamics is governed by two different types of processes that take place independently and that can be described by two different effective models. The first process is the resonant, two-photon excitation that drives the $|gg\rangle \rightarrow |ee\rangle$ transition via a second-order process, and the subsequent incoherent decay towards $|gg\rangle$, passing through the single-photon subspace.  This is described by an effective three-level cascade model, that we denote Model 2P, see Fig.~\ref{fig:fig2_effective_models}(a).

The second mechanism is the excitation of the one-photon subspace $\{|S\rangle, |A\rangle \}$ via first-order processes. This is described by a three-level Vee model that we denote Model 1P, see Fig.~\ref{fig:fig2_effective_models}(b), that excludes the doubly excited state $|ee\rangle$. 

Our key assumption will be to consider that either both types of processes take place independently and very scarcely, or that the dynamics is completely dominated by one of the processes (e.g. at the two-photon resonance condition $\Delta=0$ for the second-order processes, or the one-photon resonance conditions $\Delta   = \pm R$ for the first-order). In both cases, this means that the probabilities of occupation of excited states and coherences computed from each of these models contribute additively to the total density matrix. Adopting the notation $\rho_{i j}\equiv \langle i|\hat\rho|j\rangle$, we express this as
\begin{equation}
 \rho_{ij} = \rho_{ij}^{(1)}+\rho_{ij}^{(2)},
\label{eq:rho_relations}
\end{equation}
where $\rho_{ij}^{(2)}$  and $\rho_{ij}^{(1)}$ denote matrix elements computed from second-order processes (Model 2P) and first-order processes (Model 1P) respectively. In the following, we detail how these matrix elements are computed from each of these two approximated models.

\subsubsection{Model 2P: second-order processes}
Model 2P describes the nonlinear process of coherent, two-photon excitation by the driving laser. Therefore, it  will predict sizable probabilities for the excited states only around the two-photon resonance, $\Delta\sim 0$. For values of $\Delta$ close to zero, $|gg\rangle$ and $|ee\rangle$ form a quasi-degenerate subspace in the rotating frame of the laser,  split from the states, $|S\rangle$ and $|A\rangle$ by an energy difference $\pm R$, where we assume $R$ to be the largest energy scale in our system, so $R\gg \Delta$. The states $|gg\rangle$ and $|ee\rangle$ are not coupled to first order, $\langle ee|\hat H|gg\rangle=0$, but they are coupled through second-order processes mediated by $|S\rangle$ and $|A\rangle$, which couple to both $|gg\rangle$ and $|ee\rangle$ through the driving laser. We will assume $\Omega \ll R$ so that we can describe the effective two-photon coupling rate $\Omega_{2\mathrm p}$ between $|gg\rangle$ and $|ee\rangle$ by second-order perturbation theory, with $\hat H_\mathrm{d}$ as the perturbation:
\begin{equation}
\Omega_{2\mathrm p} = -	\sum_{i=S,A}\frac{\langle ee|\hat H_\mathrm{d}|i\rangle\langle i|\hat H_\mathrm{d}|gg\rangle}{E_i} = -\frac{2\Omega^2}{R}\cos\beta.
\label{eq:Omega_2p}
\end{equation} 
$\Omega_{2\mathrm p }$ can be understood as a two-photon Rabi frequency, featuring a quadratic scaling with $\Omega$ and a strong dependence on $\beta$,  or equivalently, the ratio $J/\delta$. One obvious consequence of Eq.~\eqref{eq:Omega_2p} is that, for $J=0$, there is not two-photon coupling between $|gg\rangle$ and $|ee\rangle$ due to destructive interference between the two possible pathways that mediate the interaction. This explains why the emergence of optical features related to the two-photon process stands as a clear evidence of coherent coupling between quantum emitters~\cite{Varada1992,Hettich2002}. The states $|ee\rangle$ and $|gg\rangle$ also experience an effective Lamb shift $\lambda$ which, notably, is the same for both of them and equal to
\begin{equation}
\lambda = \lambda_j = -	\sum_{i=S,A}\frac{\langle j|\hat H_\mathrm{d}|i\rangle\langle i|\hat H_\mathrm{d}|j\rangle}{E_i}= \Omega_{2\mathrm p }.
\label{eq:lamb-shift}
\end{equation} 
with $j=gg,ee$. 

Beyond the coherent two-photon driving, the remaining ingredient of the dynamics is the incoherent decay from $|ee\rangle$ to $|gg\rangle$, passing through one of the single-photon states, $|S\rangle$ or $|A\rangle$. Since this is an incoherent process that populates in an equal manner both single-photon states, we simplify our description by considering a single intermediate single-photon state $|1\rangle$, whose steady state population gives us the sum of the populations of $|S\rangle$ and $|A\rangle$, see Fig.~\ref{fig:fig2_effective_models}(a). The energy of this state is irrelevant in this picture since it is incoherently populated, and thus we set it to zero.

The state of the reduced subsystem $\{|gg\rangle, |ee\rangle, |1\rangle \}$ is described by a $3\times 3$ density matrix $\hat \chi^{2\mathrm p}$. Its dynamics is governed by the following master equation:
\begin{equation}
	\frac{d \hat \chi^{2\mathrm p}}{dt} = -i \left[\hat H_{2\mathrm p},\hat \chi^{2\mathrm p} \right] 
+\frac{2\gamma}{2} \mathcal{L}_{| 1 \rangle \langle ee |}\hat \chi^{2\mathrm p}+ \frac{\gamma}{2} \mathcal{L}_{|gg \rangle \langle 1 | }\hat \chi^{2\mathrm p},
\label{eq:master_eq_model1}
\end{equation}
where $\hat H_{2\mathrm p}$ is the effective two-photon Hamiltonian 
\begin{multline}
	\hat H_{2\mathrm p}= (2\Delta +\Omega_{2\mathrm p})|ee\rangle\langle ee| + \Omega_{2\mathrm p}|gg\rangle\langle gg|\\ + \Omega_{2\mathrm p}(|ee\rangle\langle gg|+|gg\rangle\langle ee|).
\label{eq:H2p}
\end{multline}
From this, we can obtain the second-order contributions to the excited-state components of the total $\hat\rho$---see Eq.~\eqref{eq:rho_relations}---establishing the following relations
\begin{subequations}
\begin{align}
\rho^{(2)}_{ee,ee} &\equiv \chi^{{2\mathrm p}}_{ee,ee}, \\
\rho^{(2)}_{S,S}  =\rho^{(2)}_{A,A} &\equiv \chi^{{2\mathrm p}}_{1,1}/2, \\
\rho^{(2)}_{A,S} &= 0.
\end{align}
\end{subequations}
Solving for the steady-state of Eq.~\eqref{eq:master_eq_model1} gives direct analytical expressions for the elements of $\hat\chi^{2\mathrm p}$, and therefore for $\hat\rho^{(2)}$. These read:
\begin{subequations}
\label{eq:rho_elements_modelI}
\begin{align}
\label{eq:rho_ee_modelI}
\rho^{(2)}_{ee,ee} &= \frac{4 \Omega^4 \cos^2 \beta }{16 \Omega^4 \cos^2 \beta+ R^2 \gamma^2 + 4R^2 \Delta^2},\\
\label{eq:rho_SS_modelI}
\rho^{(2)}_{S,S} &= \rho^{(2)}_{A,A} = \rho^{(2)}_{ee,ee}.
\end{align}
\end{subequations}
\subsubsection{Model 1P: first-order processes}
Our second model describes dynamics in which the single-photon states $|S\rangle$ and $|A\rangle$ are directly excited by the driving field through a first-order process. In essence, Model 1P consists in removing the state $|ee\rangle$ from our description, thus neglecting the two-photon excitation mechanisms that are described by Model 2P, and thus allowing only for first-order processes to occur. 
 The result is a three-level Vee system comprising the basis states $\{|gg\rangle, |A\rangle, |S\rangle \}$. The Hamiltonian in this reduced model reads
\begin{multline}
\hat H_{1\mathrm p} = (\Delta + R)|S\rangle \langle S|+(\Delta - R)|A\rangle \langle A|\\
+\Omega_A(|A\rangle\langle gg| + \mathrm{h.c.})+\Omega_S(|S\rangle\langle gg| + \mathrm{h.c.}).
\end{multline}
The driving rates $\Omega_S$ and $\Omega_A$ are simply given by $\langle S|\hat H_\mathrm{d}|gg\rangle$ and $\langle A|\hat H_\mathrm{d}|gg\rangle$ respectively, and are obtained directly from Eqs.~\eqref{eq:Hd} and \eqref{eq:eigenstates},
\begin{equation}
\Omega_{S/A}= \Omega\sqrt{1\pm\cos\beta}.\\
\label{eq:Omega_SA}
\end{equation}
In the usually considered situation of identical emitters, $\beta=0$, one finds that $\Omega_S=\sqrt 2\Omega$ and $\Omega_A = 0$, so that the antisymmetric state is dark and decoupled from the driving field. Considering now spontaneous emission in the basis of $|S\rangle$ and $|A\rangle$, we find the following master equation for the $3\times 3$ density matrix $\hat\chi^{1\mathrm p}$ of Model 1P:
\begin{multline}
\frac{d \hat\chi^{1\mathrm p}}{dt} =
		 -i \left[\hat H_{1\mathrm p},\hat\chi^{1\mathrm p}\right] +\sum_{i=S,A}\frac{\gamma_i}{2} \mathcal{L}_{{| g \rangle \langle i |}} \hat\chi^{1\mathrm p} \\
+\frac{\gamma_{C}}{2} \left( 2{| g \rangle \langle A |} \hat\chi^{1\mathrm p} {{| S \rangle \langle g |}} - \left\{{{| S \rangle }} { \langle A |}, \hat\chi^{1\mathrm p}\right\}+\text{h.c.} \right).
\label{eq:master_eq_model2}
\end{multline}
where we have defined the following decay rates
\begin{align}
	\gamma_{S/A}&=\gamma\pm\gamma_{12}\cos \beta,\\
	\gamma_C&= \gamma_{12} \sin \beta.
\end{align}
The rates $\gamma_{S/A}$ describe the standard decay of $|S/A\rangle$ towards $|gg\rangle$, while $\gamma_C$ is the rate of incoherent coupling between $|S\rangle$ and $|A\rangle$ that originates from their collective decay. Describing the one-photon dynamics in this basis allows us to easily distinguish between enhanced and suppressed channels of decay. For instance, focusing on $\gamma_A$, one can see that $\gamma_A=0$ for two close, identical emitters ($\beta=0, \gamma_{12}=\gamma$),  meaning that the antisymmetric state $|A\rangle$ gets completely decoupled from the dynamics and turns into a dark state whose population gets trapped~\cite{Fleischhauer2005,Khan2017,Zelensky2002,Jia-Hua2004}. Similarly, in that situation one finds $\gamma_S = 2\gamma$, which clearly shows the superradiant nature of the symmetric state $|S\rangle$. 

The master equation~\eqref{eq:master_eq_model2} also yields analytical solutions for the stationary state.
Once the master equation~\eqref{eq:master_eq_model2} is solved, we map the elements of the reduced density matrix $\hat\chi^{1\mathrm p}$ to the first-order contributions to the total $\hat\rho$, which we denote $\hat\rho^{(1)}$. For elements of the one-photon subspace $\{|S\rangle, |A\rangle\}$, this  map is simply given by
\begin{equation}
\rho^{(1)}_{i,j} = \chi^{1\mathrm p}_{i,j}, \quad i,j\in\{S,A\},
\end{equation}
The resulting first-order contributions to total steady-state density matrix are thus given by:
\begin{equation}
		\label{eq:rho_AA_modelII}				
{\rho^{(1)}_{{S,S}/{A,A}}}\approx \frac{4 \Omega^2 (1\pm\cos\beta)}{\gamma_{S/A}^2+4 (\Delta \pm R)^2+8 \Omega^2(1\pm\cos\beta)},
\end{equation}
and
\begin{widetext}
\begin{equation}
	\rho_{S,A}^{(1)}=\frac{2 \Omega ^2 \sin \beta  \left(\Delta ^2-R^2-2\Omega ^2\right)}{2 \left[\gamma ^2 \Delta
		^2+\left(\Delta ^2+2 \Omega ^2\right)^2\right]+\gamma ^2
		R^2 \cos (2 \beta )+R^2 \left(\gamma ^2-4 \Delta ^2+8
		\Omega ^2\right)+2 R^4-4 \Delta  R \cos \beta 
		\left(\gamma ^2+4 \Omega ^2\right)}.
\label{eq:rho_AS_modelII}
\end{equation}
\end{widetext}
The expression provided in Eq.~\eqref{eq:rho_AA_modelII} is an approximation that we obtained by eliminating $|A\rangle$ from Model 1P when calculating $\rho_{S,S}^{(1)}$, and viceversa (effectively working with two-level systems $\{|g\rangle, |S/A\rangle\}$ instead of the three-level system of Model 1P). This approximation is not necessary to obtain an analytical expression from the master equation for the three-level system in Eq.~\eqref{eq:master_eq_model2}; however the full analytical expression of $\rho_{S,A}^{(1)}$ is long and cumbersome to write, and it is very well matched by the more compact and tractable expressions in Eq.~\eqref{eq:rho_AA_modelII}.

\begin{figure*}[t]
	\begin{center}
		\includegraphics[width=0.95\textwidth]{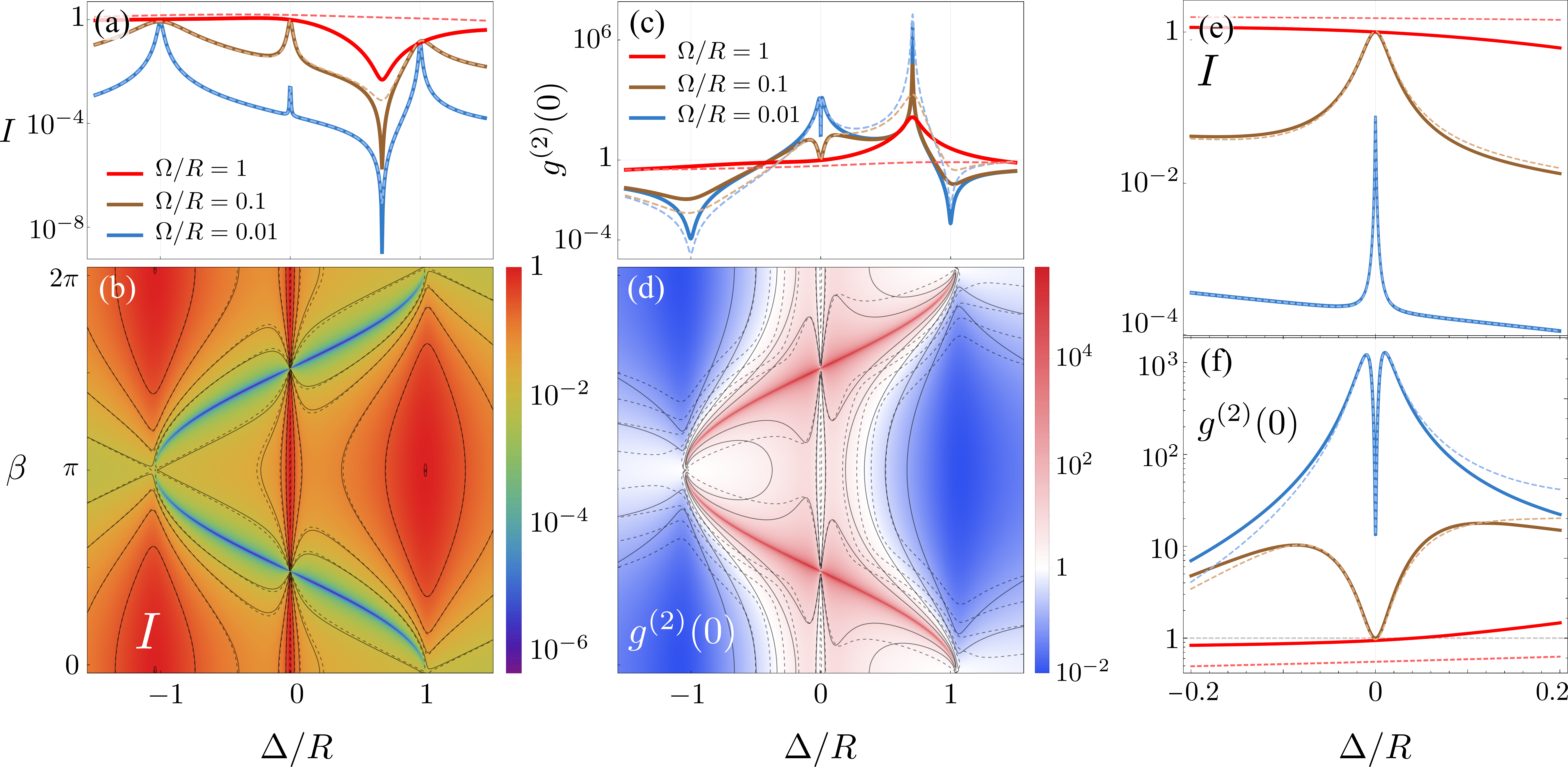}
	\end{center}
	\caption{Steady state observables of the emitted light. In all panels, dashed lines represent analytical results, solid lines are exact, numerical results.
	 (a-b) Intensity versus laser detuning (a) and both laser detuning and mixing angle $\beta$ (b). We observe a very good agreement between analytical and numerical results.
	 (c-d) Second-order correlation function at zero delay versus laser detuning (c) and both laser detuning and mixing angle $\beta$ (d). Best agreement between analytical and numerical results is found around $\Delta\sim 0$.
(e-f) provide a zoom of panels (a) and (c) respectively, around $\Delta \sim 0$. This zoom highlights the good agreement between numerical and analytical results. Parameters:  $\gamma/R=10^{-3}$, $\gamma_{12}=0.999\gamma$. In panels (a), (c), (e), and (f), $\beta=\pi/4$. In panels (b) and (d), $\Omega/R=0.1$.}
	\label{fig:fig3-observables}
\end{figure*}

In principle, Model 1P ignores the excitation of the two-photon subspace $|ee\rangle$ via successive, first-order excitation processes. This is justified since, focusing around the region we are most interested in, the two-photon resonance $\Delta\approx 0$, the contribution of this mechanism to $\rho_{ee}$ is rather small compared to that of second-order processes. However, when one approaches the limit $\beta\approx \pi/2$, the two-photon population originated by second-order processes, $\rho_{ee,ee}^{(2)}$, tends to zero, as can be seen from Eq.~\eqref{eq:rho_ee_modelI}, and the small contribution from first-order processes may dominate and become relevant. In that particular regime, we can infer the value of the first-order contribution $\rho^{(1)}_{ee,ee}$ even if the state $|ee\rangle$ is not included in Model 1P. This can be done by noting that, for $\beta\sim \pi/2$, the two quantum emitters are essentially decoupled, and the population $\rho_{ee,ee}$ stems from the simultaneous but independent excitation of both emitters. This yields factorizable correlations of the type $\langle\hat\sigma_1^+ \hat\sigma_2^+\hat\sigma_1\hat\sigma_2\rangle =\langle\hat\sigma_1^+\hat\sigma_1\rangle \langle\hat\sigma_2^+\hat\sigma_2\rangle $. Since $\rho_{ee,ee} = \langle\hat\sigma_1^+ \hat\sigma_2^+\hat\sigma_1\hat\sigma_2\rangle$, this factorization allows us to use the expressions of the populations obtained from Model 1P, $\langle\hat\sigma_i^+\hat\sigma_i\rangle^{(1)}$, to estimate first-order contributions to the occupation $\rho^{(1)}_{ee,ee}=\langle\hat\sigma_2^+\hat\sigma_2\rangle^{(1)}\langle\hat\sigma_1^+\hat\sigma_1\rangle^{(1)}$, even if $|ee\rangle$ was not explicitly included in the model. For $\beta\sim \pi/2$, we have that $\langle\hat\sigma_2^+\hat\sigma_2\rangle\langle\hat\sigma_1^+\hat\sigma_1\rangle\approx \rho_{S,S}\rho_{A,A}$. Therefore, we will define
\begin{equation}
\rho_{ee,ee}^{(1)}\equiv \rho_{ee,ee}^{(1)}(\Delta\sim 0, \beta\sim \pi/2) = \rho_{S,S}^{(1)}\rho_{A,A}^{(1)}.
\label{eq:rho_ee_modelII}			
\end{equation}
This expression is necessary to regularize the expected two-photon population $\rho_{ee,ee}$ in the limit of uncoupled emitters, $\beta=\pi/2$, and basically unimportant in any other case.

\section{Steady state observables of the emitted light}
\label{sec:2_steady_state}

Following the scheme of  Eq.~\eqref{eq:rho_relations}, we can now combine the results provided by the effective models just discussed  and obtain an estimate of the total steady-state density matrix $\hat\rho$. The approximations used are expected to hold particularly well around the region of interest $\Delta\sim 0$, i.e. the two-photon resonance. In particular, the most relevant density matrix elements for subsequent calculations read:
\begin{subequations}
\begin{align}
\label{eq:rho_ee}
\rho_{ee,ee} &\approx \rho_{ee,ee}^{(1)}+ \rho_{ee,ee}^{(2)},\\
\label{eq:rho_SS}
\rho_{S,S} &= \rho_{S,S}^{(1)}+ \rho_{S,S}^{(2)},\\
\label{eq:rho_AA}
\rho_{A,A} &= \rho_{A,A}^{(1)}+ \rho_{A,A}^{(2)},\\
\label{eq:rho_AS}
\rho_{S,A} &= \rho_{S,A}^{(1)}.
\end{align}
\end{subequations} 
These density matrix elements allow us to obtain steady-state observables of the fluorescent light emitted by the system, which is done by establishing a connection between the radiated electric field operator and the annihilation operators of the quantum emitters. For instance, if we consider the emitters to be located at the origin and to have the same dipole moment $\bm\mu$ (as we do throughout this work), the positive frequency part of far-zone electric field operator radiated by the two quantum emitters is given by $\hat{\mathbf{E}}^{(+)}(\mathbf{r},t)= \hat{{E}}^{(+)}(\mathbf{r},t)\mathbf{u_x}$, where $\mathbf{u_x}$ is a unit vector perpendicular to $\mathbf r$ and contained within the plane spanned by $\bm \mu$ and $\mathbf r$~\cite{Scully1997}, and 
\begin{equation}
\hat{{E}}^{(+)}(\mathbf{r},t)=\frac{\omega_0^2}{4\pi\epsilon_0 c^2 |\mathbf{r}|^2}|\bm{\mu}\times\mathbf{r}|( \hat\sigma_1 +  \hat\sigma_2)(t).
\end{equation}
Other detection schemes can give a different spatial distribution of the field,  e.g. imaging by focusing the field with a lens yields a distribution given by its point spread function. Nevertheless,  the relationship
\begin{equation}
\hat{{E}}^{(+)}(\mathbf{r},t)\propto ( \hat\sigma_1 +  \hat\sigma_2)(t)
\label{eq:E-propto-sigma}
\end{equation}
 will hold provided the dipole moments are equal and their separation is small compared to the resonant wavelength of the field $k r_{12}\ll 1$~\cite{Novotny2009}. 
This regime is of particular interest for our work, since a good understanding of the system dynamics and quantum optical properties of the emission becomes essential to infer inter-emitter distances  below the diffraction limit~\cite{Hettich2002}, which has important applications for microscopy and superresolution imaging~\cite{Schwartz2012}. Therefore, we will assume the proportionality relation \eqref{eq:E-propto-sigma} throughout the text.

The two main observables of the flourescent emission  that we will analyze here are  mean intensity and second-order correlation function.

\subsection{Mean intensity}
From the relation~\eqref{eq:E-propto-sigma}, we see that the mean intensity of the signal is given by ${\langle \hat{{E}}^{(-)}\hat{{E}}^{(+)}}\rangle \propto \langle \hat I\rangle$, where we have defined the intensity operator
\begin{equation}
\hat I  \equiv  (\hat\sigma_1^+ + \hat\sigma_2^+)(\hat\sigma_1 + \hat\sigma_2).
\label{eq:intensity_operator}
\end{equation}
The steady-state mean value of the intensity operator $I\equiv \langle \hat I\rangle $ can be expressed in terms of the density matrix elements as
\begin{multline}
	I=2\rho_{ee,ee}+\rho_{S,S}+\rho_{A,A}\\
	+\cos \beta (\rho_{S,S}-\rho_{A,A}) 	+ 2 \sin \beta \mathrm{Re}\left[ \rho_{S,A} \right] .
\label{eq:intensity_matrix_elements}
\end{multline}
Equation~\eqref{eq:intensity_matrix_elements}, together with \cref{eq:rho_ee_modelI,eq:rho_SS_modelI},~\cref{eq:rho_AS_modelII,eq:rho_AA_modelII,eq:rho_ee_modelII} and~\cref{eq:rho_ee,eq:rho_SS,eq:rho_AA,eq:rho_AS}, provides a direct analytical expression for $I$. 

Our analytical results are shown in comparison with numerical results in Fig.~\ref{fig:fig3-observables}. Panel (a) depicts the intensity $I$ versus the laser detuning $\Delta$; (b) shows $I$ versus both $\Delta$ and $\beta$, and  (e) represents a zoom of (a) around the two-photon excitation regime, $\Delta\sim 0$. There are three characteristic high-intensity peaks corresponding to the values $\Delta =\{-R,0,R\}$. The two peaks at $\pm R$ correspond to the resonant excitation of the states $|A/S\rangle$ and has therefore a first-order origin described by Model 1P. Their relative height depends strongly on $\beta$ (i.e. on the ratio between the dipole-dipole coupling and the relative detuning between the emitters), as clearly seen in Fig.~\ref{fig:fig3-observables}(b). Model 1P provides a good match with the exact, numerical calculation provided that the occupation probabilities are small enough for the simultaneous excitation probability to be negligible.  When $\beta=0$, i.e. the usually considered situation of resonant (identical) emitters, only the symmetric state $|S\rangle$ gets significantly populated at $\Delta = -R$, while the resonance with the $|A\rangle$ state at $\Delta=R$ is suppressed. The reverse situation is found for $\beta=\pi/2$, where both emitters are uncoupled and thus only classical correlations between them can exist. In this situation, the curve is symmetric around $\Delta=0$ and the two peaks at $\Delta=\pm R$, which in this case correspond to the resonant driving of each of the independent QEs, have similar values.

\begin{figure}[t]
	\begin{center}
		\includegraphics[width=0.9\columnwidth]{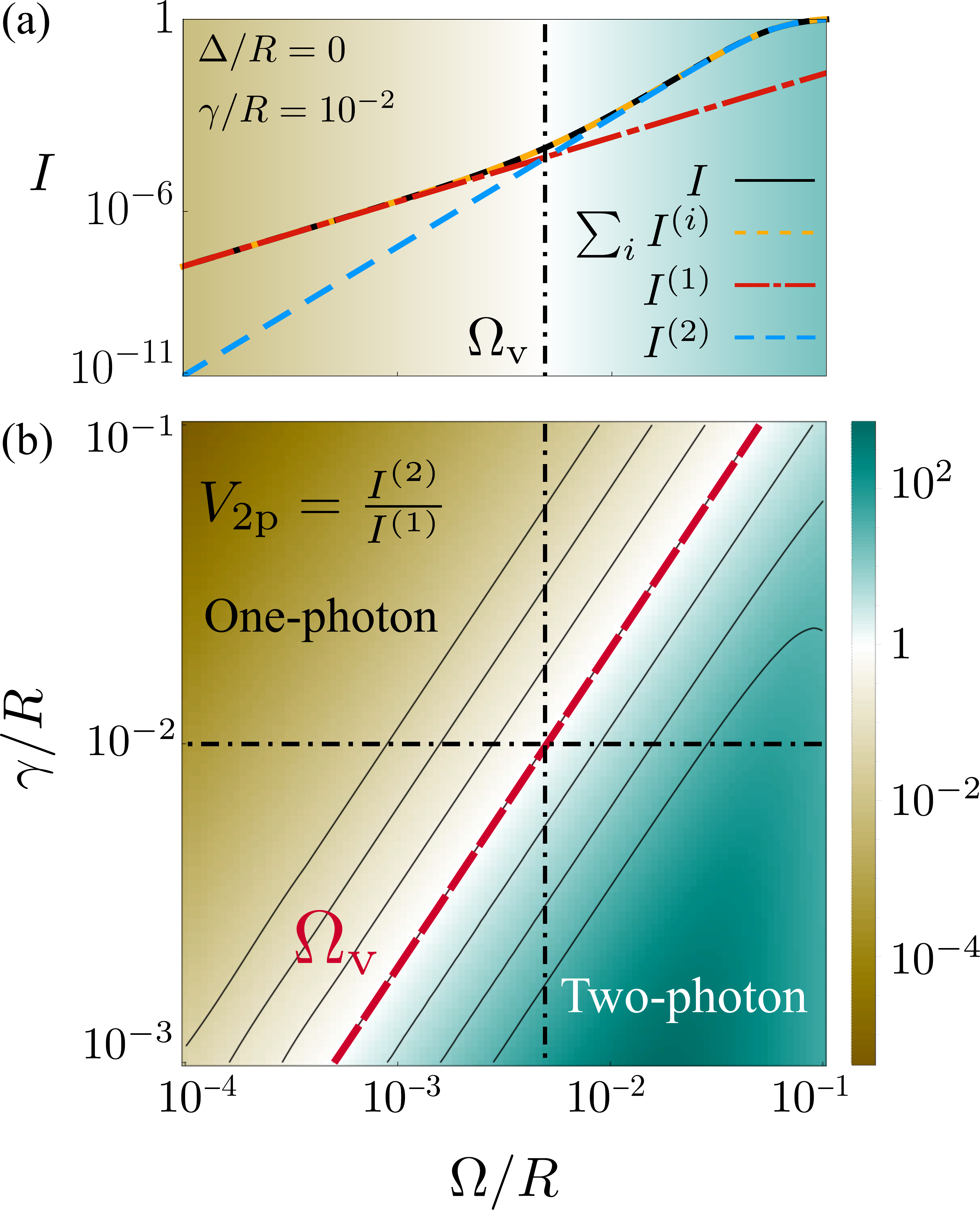}
	\end{center}
	\caption{Regimes of visibility of two-photon physics. (a) Intensity of emission $I$ computed numerically versus driving amplitude at the two-photon resonance $\Delta=0$, together with first- and second-order contributions, $I^{(1)}$ and $I^{(2)}$ obtained analytically. Their sum recovers the exact, numerical value of $I$. $\Omega_\mathrm{v}$ marks the two-photon visibility crossover where second-order contributions are larger than first-order ones, $I^{(2)}>I^{(1)}$. $\gamma/R=10^{-2}$. (b) Two photon visibility $V_{2\mathrm p}=I^{(2)}/I^{(1)}$ versus $\gamma$ and $\Omega$. Dashed-red line marks the analytical value of $\Omega_\mathrm{v}$ given by Eq.~\eqref{eq:Omega_visibility}, matching perfectly the contour line $V_{2\mathrm p}=1$. Parameters: $\beta=\pi/4$, $\gamma_{12}=0.999\gamma$.}
	\label{fig:fig4-visibility}
\end{figure}

The central peak at $\Delta=0$ is arguably the most relevant feature given its implications for microscopy and imaging~\cite{Hettich2002}. This peak emerges from the resonant two-photon excitation enabled by the coherent coupling between emitters, and thus is fully described by the contributions of second-order processes from Model 2P,~\cref{eq:rho_ee_modelI,eq:rho_SS_modelI}. Our analytical solution allows us to establish that both the height and width of this peak scale as $\Omega^4 \cos^2\beta$. As expected, the peak vanishes for uncoupled emitters ($\beta=\pi/2$) as a consequence of the destructive interference of the two-excitation pathways, which yield $\Omega_{2\mathrm p}=0$. 
Our analytical expression of $\rho_{ee,ee}$ \eqref{eq:rho_ee_modelI} gives us the possibility to use the intensity of the two-photon peak at $\Delta=0$ to infer the value of $\beta$. On the other hand, $R$ can also be easily estimated from the position of the one-photon peaks. The knowledge of these two magnitudes can then be combined  to obtain information about the internal structure of the quantum emitters, i.e., their natural energy detuning $2\delta$ and coherent coupling rate $J$. In turn, this allows one to infer quantities such as the inter-emitter distance $k r_{12}$, highly relevant for technological applications like superresolution imaging~\cite{Hettich2002}. 

Given its importance, it is desirable to determine the set of conditions under which the two-photon peak will be visible. Following our approach of separating contributions from first- and second-order processes, the total intensity can be written as $I = I^{(1)} + I^{(2)}$. The two-photon peak arises from the resonant contribution $I^{(2)}$, while the off-resonant, first-order contribution $I^{(1)}$ gives a featureless background at $\Delta=0$ which, under certain conditions, can be brighter than the two-photon peak and hide it. In particular, since the population of second-order origin $\rho_{ee,ee}^{(2)}$ responsible for the two-photon peak have quartic scaling with $\Omega$, while first-order processes yield populations that scale quadratically with $\Omega$, there must be a value $\Omega_{\mathrm v}$ below which first-order processes dominate, see Fig.~\ref{fig:fig4-visibility}(a). To determine $\Omega_\mathrm{v}$, we can define a two-photon visibility $V_{2\mathrm p}$ as the ratio $V_{2\mathrm p}= I^{(2)}/I^{(1)}$, such that the two-photon peak will be visible when $V_{2\mathrm p}>1$. $I^{(2)}$ and $I^{(1)}$ can be computed from the first- and second-order contributions to $\rho_{ee,ee}$, $\rho_{S,S}$, $\rho_{A,A}$ and $\rho_{S,A}$, using \cref{eq:rho_ee_modelI,eq:rho_SS_modelI} and \cref{eq:rho_ee_modelII,eq:rho_AA_modelII,eq:rho_AS_modelII} respectively. Introducing these into the equation $V_{2\mathrm p}=1$ and solving it, we obtain a value for the minimum necessary driving amplitude $\Omega_\mathrm v$ that guarantees two-photon visibility. The equation is greatly simplified in the regime that we consider in this paper, $\Omega, \gamma \ll R $, yielding 
\begin{equation}
\Omega_\mathrm{v} \approx R \sqrt{\frac{2}{\tan ^2 \beta +8R^2/\gamma ^2}} \approx \frac{\gamma}{2},
\label{eq:Omega_visibility}
\end{equation}
where the last approximation applies provided $\gamma \ll R\tan\beta$. These results are summarized and confirmed by exact numerical calculations in Fig.~\ref{fig:fig4-visibility}. In panel (a), we show that the sum of our analytical estimations of $I^{(1)}$ and $I^{(2)}$ recovers the exact value of $I$ computed numerically, which, as discussed above, features a transition from a $\propto \Omega^2$ to a $\propto \Omega^4$ scaling at $\Omega_\mathrm{v}$, which marks the onset of visibility of the two-photon peak, i.e., the emergence of features characteristic of the two-photon dynamics. The two-photon visibility $V_{2\mathrm p}$ is shown in the full $(\gamma,\Omega)$ space in panel (b), where the approximated expression for $\Omega_\mathrm{v}$ provided in Eq.~\eqref{eq:Omega_visibility} is shown to match perfectly the condition $V_{2\mathrm p} = 1$. 

To end our discussion on the mean intensity of the radiated field, we observe that a destructive interference also appears  for values $\beta\neq\pi/2$. This  can be seen in Fig.~\ref{fig:fig3-observables}(b), where a destructive interference dip manifests when the laser detuning is $\Delta=R\cos\beta$. However, unlike the particular case $\beta=\pi/2$, this is not a destructive quantum interference between excitation pathways in the internal system dynamics, but an optical one, taking place in the radiated electric field and well described by the interference terms appearing in Eq.~\eqref{eq:intensity_matrix_elements}. Indeed, using our Model 1P, we find that the dip is given by point where all the one-photon subspace terms proportional to $\rho_{S,S}$, $\rho_{A,A}$ and $\rho_{S,A}$ in Eq.~\eqref{eq:intensity_matrix_elements} add up to zero. As we discuss below, the small amount of remaining light retains a very strong two-photon character.

\subsection{Second-order correlation function}
The zero-delay second-order correlation function is defined as $g^{(2)}(0) = {\langle {{\hat E}}^{(-)} {{\hat E}}^{(-)} {{\hat E}}^{(+)}}  {{\hat E}}^{(+)} \rangle /  \langle {{\hat E}}^{(-)} {{\hat E}}^{(+)}\rangle^2$, and thus it can be written as
\begin{equation}
	g^{(2)}(0)= \frac{\langle ({\hat\sigma_1}^+ + {\hat\sigma_2}^+)^2 ({\hat\sigma_1} + {\hat\sigma_2})^2  \rangle}{\langle ({\hat\sigma_1}^+ + {\hat\sigma_2}^+)    ({\hat\sigma_1} + {\hat\sigma_2}) \rangle^2 } = \frac{4 \rho_{ee,ee}}{{I}^2}.
\label{eq:g2}
\end{equation}
This value quantifies the probability of detecting two photons simultaneously, normalized by the probability of doing so in a classical coherent field of similar intensity. As seen in Eq.~\eqref{eq:g2}, in our case this is directly related to the probability of occupying the doubly excited state $|ee\rangle$. The numerical and analytical results are summarized in Fig.~\ref{fig:fig3-observables}(c-d). 
Our approximated analytical methods are able to describe accurately the population of the two-photon subspace---and therefore $g^{(2)}(0)$---around the two-photon resonance $\Delta\sim 0$, provided $\Omega \ll R$. This can be seen more clearly in the zoom around the two-photon resonance depicted in Fig.~\ref{fig:fig3-observables}(f). Away from the two-photon resonance, the population of the doubly excited state becomes much smaller and it is established by a more complicated mixture of one-photon and two-photon processes that our approximated models fail to capture. These small occupations of $|ee\rangle$, nevertheless, contribute very little to the actual intensity of radiation emitted, which away from $\Delta\sim 0$ is dominated by the one-photon subspace and thus is well described by our models, c.f. Fig.~\ref{fig:fig3-observables}(b). Interestingly, the maximum values of $g^{(2)}(0)$ are found at the dip of destructive interference discussed before, where the first-order contributions to the total emission interfere destructively and thus the small amount of remaining emission stems mainly from second-order processes, yielding a strong probability of detecting two photons. 

At the two-photon resonance $\Delta=0$, the value of the $g^{(2)}(0)$ is sharply reduced and tends to 1 from above as $\Omega$ increases, c.f. Fig.~\ref{fig:fig3-observables}(f). The reason for this is that, as $\Omega$ increases, the light emitted is less coherent ($\langle \hat\sigma_i\rangle\rightarrow 0$), and thus the emission converges to that of two incoherent quantum emitters~\cite{Sanchez2020}. The dip of $g^{(2)}$ precisely at $\Delta=0$ is a typical feature of multi-photon processes at resonance~\cite{SanchezMunoz2014}, and it is simply a consequence of the increased intensity of emission.

\section{Spectrum of Two-photon Resonance Fluorescence}
\label{sec:3-spectrum}

\begin{figure*}[t!]
	\begin{center}
		\includegraphics[width=0.95\textwidth]{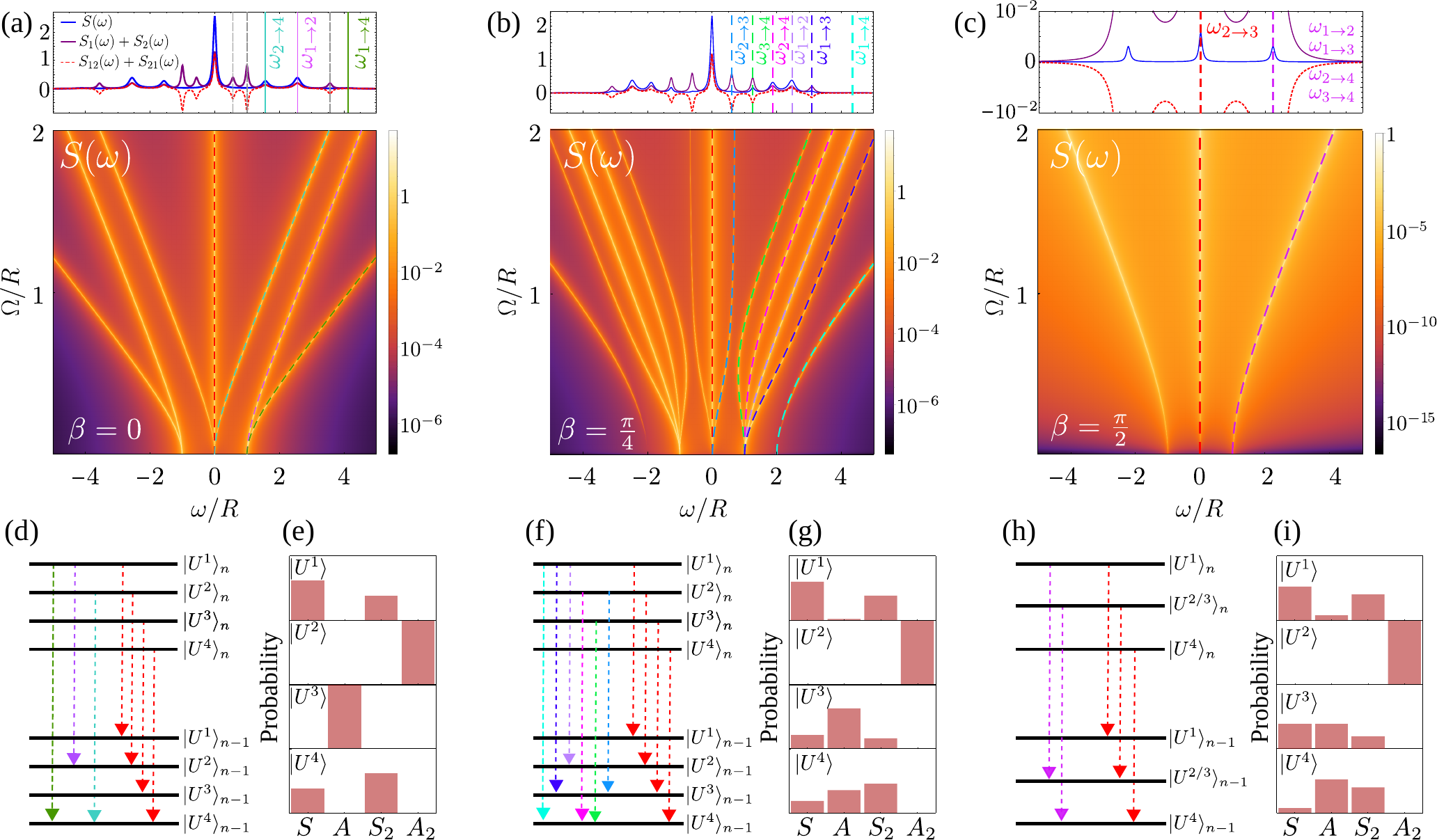}
	\end{center}
	\caption{Resonance fluorescence spectrum at the two-photon resonance $\Delta=0$, for $\beta=0$ (a), $\beta=\pi/4$ (b) and $\beta=\pi/2$ (c). In (a-c), upper plots depict the spectrum at $\Omega/R=1$ (solid blue lines) and the different contributions in its expansion in Eq.~\eqref{eq:spectrum_expansion} (solid purple and dashed lines) , allowing to see cases in which peaks are not visible due to destructive interference, but visible if only the emission from a single emitter is collected. The lower panels in (a-c) show the emergence of sidebands as $\Omega/R$ increases; in the perturbative regime $\Omega\ll R$, these correspond to transitions between two-photon dressed states. Positive-frequency peaks are marked with dashed lines, which are identified with the corresponding transition between the dressed-atom ladder of eigenstates in panels (d,f,g,i). Panels (e,g,i) depict the structure of the eigenstates at $\Omega/R=1$, in the basis $\{|S\rangle, |A_2\rangle, |S_2\rangle, |A\rangle \}$. Parameters: in upper plots of (a-c), $\gamma/R = 0.1$. In the lower panels of (a-c), $\gamma/R = 10^{-3}$, $\gamma_{12}=0.999\gamma$}
	\label{fig:fig5-spectrum-omega}
\end{figure*}

The fluorescence spectrum of the emitted radiation yields useful information about the energy transitions that can take place among the dressed states of the hybridized light-matter system. The most celebrated example of the revealing character of this type of measurement is the Mollow triplet spectrum in the emission from a two-level atom~\cite{Mollow1969}. Its characteristic lineshape with three peaks provides key information about the structure of dressed energy levels~\cite{CohenTannoudji1998}, and it serves a source of strongly correlated non-classical light~\cite{Gonzalez-Tudela2013a,Ulhaq2012,sanchezmunoz14b,Peiris2015,Peiris2017,LopezCarreno2017}. 

The resonance fluorescence spectrum of two interacting quantum emitters displays a more complex structure than the Mollow triplet, which has been reported for the case of identical emitters~\cite{Ficek1983,Darsheshdar2021}, and for the analogous case of a coherently driven three-level system at the two-photon resonance~\cite{Gasparinetti2019}. The resulting spectrum has a seven-peaked structure with a central peak and six sidebands. 
Here, we recover this same result for the case of identical emitters~[$\beta=0$, see Fig.~\ref{fig:fig5-spectrum-omega}(a)], but extend it for non-indentical emitters ($\beta>0$), where we show that the spectrum develops an even more complicated structure with up to 13 peaks [see Fig.~\ref{fig:fig5-spectrum-omega}(b)], that end up converging to the three peaks characteristic of the Mollow triplet for completely uncoupled emitters, $\beta=\pi/2$ [see Fig.~\ref{fig:fig5-spectrum-omega}(c)]. In the perturbative regime $\Omega \ll R$, we are able to describe the position of these peaks from the hybridization between the quantum emitters and photon pairs using the dressed energy levels described by Model 2P. The understanding and effective Hamiltonian obtained from this effective model allows one to describe analytically the location and origin of the spectral resonances in the perturbative regime $\Omega \ll R$, and also to obtain insights about this structure beyond that regime, when $\Omega > R$, which we explore numerically. 

The spectrum of emission is given by the Fourier transform of the two-time correlation function of the radiated electric field $\langle \hat E^{(-)} (t) \hat E^{(+)}(t+\tau) \rangle$. For convenience, we will define a general spectral function
\begin{equation}
S(\omega;\hat A,\hat B) = \lim_{t\rightarrow \infty}\frac{1}{\pi} \Re\int_0^\infty d\tau\, e^{i\omega \tau} \langle \hat A (t)\\  \hat B(t+\tau) \rangle.
\label{eq:general-spectral-function}
\end{equation}
By using the quantum regression theorem to express the two-time correlation function~\cite{Carmichael1993} and formally integrating Eq.~\eqref{eq:general-spectral-function}, we can write $S(\omega;\hat A, \hat B)$ in a computationally convenient form~\cite{sanchezmunoz2019a},
\begin{equation}
S(\omega;\hat A, \hat B)=\frac{1}{\pi} \text{Re} \text{Tr} \left\{ - \hat B \frac{1}{\mathcal{L}+i\omega } \left[\hat \rho \left( \hat A - \mathrm{Tr}[\hat\rho \hat A]\right) \right]\right\},
\label{eq:spectrum-equation}
\end{equation}
where $\hat\rho$ is the steady-state density matrix, and $\mathcal L$ is the Liouvillian superoperator that represents the master equation~\cite{Breuer2007}, where $\hat\rho$ has a vectorial form, $\partial_t|\hat \rho\rangle = \mathcal L |\hat\rho\rangle$. Considering the relation Eq.~\eqref{eq:E-propto-sigma} between the radiated field and the raising/lowering operators of the quantum emitters, and disregarding global factors, we can express the spectrum of resonance fluorescence as 
\begin{equation}
S(\omega) =S(\omega; \hat\sigma_1^+ + \hat\sigma_2^+, \hat\sigma_1 + \hat\sigma_2).
\label{eq:spectrum_total}
\end{equation}
The spectra computed as a function of $\Omega$ for different values of $\beta$ are shown in Fig.~\ref{fig:fig5-spectrum-omega}, for a laser detuning at the two-photon resonance, $\Delta=0$.

\emph{Perturbative regime}---The position of all the peaks in the spectra can be obtained from the possible transitions among energy levels in the system Hamiltonian. Within the perturbative regime $\Omega \ll R$, at $\Delta=0$ the ground and excited states $|gg\rangle$ and $|ee\rangle$ are resonantly coupled by the two-photon excitation described by the Hamiltonian in Eq.~\eqref{eq:H2p}: these two states then hybridize into a symmetric and antisymmetric states that we denote $|S_2\rangle$ and $|A_2\rangle$, defined as
\begin{equation}
| S_2/A_2 \rangle \equiv \frac{1}{\sqrt{2}}\left( |gg \rangle \pm  | ee \rangle \right).
\end{equation} 
The resulting set of eigenstates is given by $\{|S\rangle, |A_2\rangle, |S_2\rangle, |A\rangle \}$, ordered by decreasing energy, and their correspondent eigenenergies in the rotating frame of the laser are:
\begin{align}
&E_1 = E_{S} = R + 2(1+\cos\beta)\Omega^2/R,\\
&E_2 = E_{A_2} = 0,\\
&E_3 = E_{S_2} = -4\Omega^2 \cos\beta/R,\\
&E_4 = E_{A} = -R-2(1-\cos\beta)\Omega^2/R.
\end{align}
In the same way that the eigenenergies of $A_{2}$ and $S_{2}$ take into account the Lamb shifts induced by the coupling to $|A\rangle$ and $|S\rangle$---see Eq.~\eqref{eq:lamb-shift}---, $E_{A}$ and $E_{S}$ also include the Lamb shifts of states $|S\rangle$ and $|A\rangle$ due to their coupling to $|ee\rangle$ and $|gg\rangle$,  given by
\begin{equation}
\lambda_j = -\sum_{i=ee,gg}\frac{\langle j|\hat H_\mathrm{d}|i\rangle\langle|i\hat H_\mathrm{d}|j\rangle}{E_i-E_j}
\end{equation} 
with $j=S,A$. This gives $\lambda_{S/A}=\pm 2(1-\cos\beta)\Omega^2/R$.

The energy differences between these eigenvalues give us the transition frequencies that can observed as distinct peaks in the spectrum. Defining the transition energies $\omega_{i\rightarrow j} \equiv E_i - E_j$, the position of the six positive-frequency sidebands with respect to the laser frequency are defined, in the perturbative regime, by the following equations:
\begin{align}
\label{eq:transition_omega_1}
\omega_1 &= \omega_{1\rightarrow 4} = 2R+\frac{4\Omega^2}{R},\\
\label{eq:transition_omega_2}
\omega_2 &= \omega_{1\rightarrow 3} = R + \frac{2\Omega^2(3\cos\beta+1)}{R},\\
\label{eq:transition_omega_3}
\omega_3 &= \omega_{1\rightarrow 2} = R + \frac{2\Omega^2(\cos\beta+1)}{R},\\
\label{eq:transition_omega_4}
\omega_4 &= \omega_{2\rightarrow 4} = R - \frac{2\Omega^2(\cos\beta-1)}{R},\\
\label{eq:transition_omega_5}
\omega_5 &= \omega_{3\rightarrow 4} = R - \frac{2\Omega^2(3\cos\beta-1)}{R},\\
\label{eq:transition_omega_6}
\omega_6 &= \omega_{2\rightarrow 3} = \frac{4\Omega^2\cos\beta}{R}.
\end{align}
The sidebands at negative frequencies come from the reversed processes outline above, $\omega_{i\rightarrow j} = -\omega_{j\rightarrow i}$. The central peak at $\omega_0 = 0$ is given by transitions between similar states, $\omega_{i\rightarrow i}$. These equations provide the position of all the possible 13 peaks that can be observed in the fluorescence spectrum within the perturbative regime $\Omega \ll R$. 

The six positive sidebands in \cref{eq:transition_omega_1,eq:transition_omega_2,eq:transition_omega_3,eq:transition_omega_4,eq:transition_omega_5,eq:transition_omega_6}, however, are not visible for all values of $\beta$. The case of coupled identical emitters $\beta=0$ features only three positive sidebands, yielding a total of seven peaks as has been noted before~\cite{Ficek1983,Darsheshdar2021}. In particular, the peaks that are not visible are those that involve transitions that start or end at the one-photon antisymmetric state $|A\rangle$, which for the perturbative regime correspond to the peaks $\omega_2$, $\omega_5$ and $\omega_6$ in \cref{eq:transition_omega_1,eq:transition_omega_2,eq:transition_omega_3,eq:transition_omega_4,eq:transition_omega_5,eq:transition_omega_6}. The reason why these peaks are not visible is due to the destructive interference phenomena that takes place from the equal contribution from both quantum emitters to the radiated electric field, c.f. Eq.~\eqref{eq:E-propto-sigma}. This can be seen in Fig.~\ref{fig:fig5-spectrum-omega}(a). Indeed, expanding the expression of the total spectrum of emission, Eq.~\eqref{eq:spectrum_total}, we obtain:
\begin{equation}
S(\omega ) = S_1(\omega) + S_2(\omega) + S_{12}(\omega) + S_{21}(\omega)
\label{eq:spectrum_expansion}
\end{equation}
where $S_{ij}(\omega) \equiv S(\omega; \hat\sigma_i^+, \hat\sigma_j)$ and $S_i(\omega) \equiv S_{ii}(\omega)$. $S_1(\omega)$ and $S_2(\omega)$ describe the spectrum of emission that would be obtained by detecting only the radiation emitted by the QE 1 and 2, respectively. On the other hand, $S_{12}(\omega)$ and $S_{21}(\omega)$ are interference terms arising from the superposition of both fields. In Fig.~\ref{fig:fig5-spectrum-omega}(a), we show that the missing peaks for $\beta=0$ (all involving the state $|A\rangle$) are indeed visible if $S_1(\omega)$ or $S_2(\omega)$ are measured independently (e.g. if their emission is collected locally), but they interfere destructively if the fields emitted by both QEs are superimposed, explaining the absence of these peaks in the total spectrum.
For $\beta>0$, this perfect destructive interference does not occur, and all the possible 13 peaks are visible in the  spectrum of emission. Finally, the opposite limit of completely decoupled non-identical emitters, $\beta=\pi/2$, yields a three-peaked structure which corresponds to the Mollow triplet of emission from the two independent emitters. In this limit, many of the transitions in \cref{eq:transition_omega_1,eq:transition_omega_2,eq:transition_omega_3,eq:transition_omega_4,eq:transition_omega_5,eq:transition_omega_6}, become degenerate, so that only three peaks can be seen. The reason for having fewer peaks here is obviously different from the destructive interference seen at $\beta=0$, since it responds to the internal structure of transition energies available within the dressed energy levels.

Away from $\beta=\pi/2$, the frequencies in \cref{eq:transition_omega_1,eq:transition_omega_2,eq:transition_omega_3,eq:transition_omega_4,eq:transition_omega_5,eq:transition_omega_6}
describe transitions between dressed light-matter states in which the emitters are hybridized with photon pairs. This strong two-photon character is evidenced by the quadratic scaling of these frequencies with $\Omega$, instead of the linear scaling with $\Omega$ that one finds, e.g., for the position of the sidebands in the standard Mollow triplet . For this reason, we refer to the sidebands described by \cref{eq:transition_omega_1,eq:transition_omega_2,eq:transition_omega_3,eq:transition_omega_4,eq:transition_omega_5,eq:transition_omega_6} as \emph{two-photon sidebands}. In order to be able to observe two-photon sidebands, we need the energy separation $\Delta E = \omega_2-\omega_3 = \omega_4-\omega_5 = 2\Omega_{2\mathrm p}$ to be larger than the decay rate of spontaneous emission, $\gamma$. The condition $2\Omega_{2\mathrm p}>\gamma$ yields the following equation for the two-photon saturation amplitude $\Omega_{2\mathrm{PS}}$ that marks the onset of the resolved two-photon sideband regime:
\begin{equation}
\Omega_{2\mathrm{PS}} = \frac{1}{2}\sqrt{\frac{R\gamma}{\cos\beta}}.
\label{eq:Omega_2PS}
\end{equation}
The previous expression tells us that $\Omega_{\mathrm 2PS}\ll R$,  meaning that the two-photon sidebands can be developed within the perturbative regime, provided that the Rabi splitting is large so that $R\gg\gamma$, and that $\beta$ is not too close to $\pi/2$. As $\beta$ approaches $\pi/2$, larger values of $\Omega$ are necessary to resolve the two-photon sidebands. When values $\Omega_{2\mathrm{PS}}\sim R$ are reached, the perturbative approach used below does not apply, meaning that sidebands developed purely from two-photon hybridization are no longer observable since first-order processes dominate before the former are visible. Setting the condition $\Omega_{2\mathrm{PS}}\approx R$, we find that this would occur approximately at a mixing angle $\beta_\mathrm{max} \approx \arccos(\gamma/4R)$. One can see, however, that for the typical values considered in this text, i.e. $\gamma = 10^{-3} R$, $\beta_\mathrm{max} \approx 0.9998 \pi/2$, i.e. two-photon sidebands should be visible for most of the range $\beta \in [0,\pi/2]$. As we will see, the two-photon saturation amplitude $\Omega_{2\mathrm{PS}}$ has a great importance for metrological applications, since the onset of the resolved two-photon sidebands regime marks the point of maximum sensitivity for optical estimations of the inter-emitter distance. 

\begin{figure*}[t]
	\begin{center}
		\includegraphics[width=0.95\textwidth]{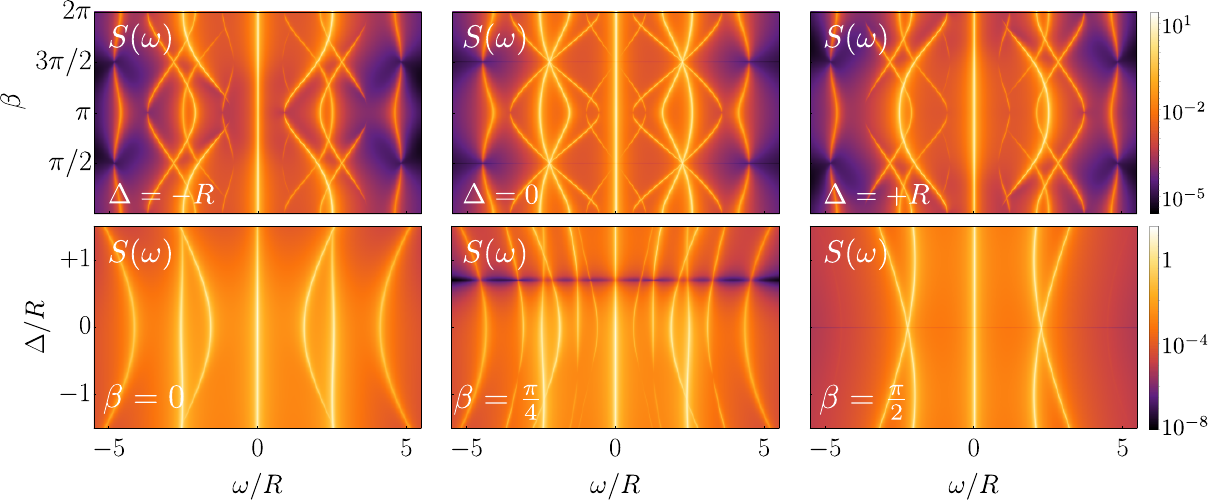}
	\end{center}
	\caption{Spectrum of resonance fluorescence versus $\beta$ (upper panels) and $\Delta$ (lower panels). Upper panels: from left to right, $\Delta=\{-R,0,R\}$.  Lower panels: from left to right, $\beta = \{0,\pi/4,\pi/2 \}$.
		  Parameters: $\gamma/R=10^{-2}$, $\Omega/R = 1$, $\gamma_{12}=0.999\gamma$. }
	\label{fig:fig6-spectrum-Delta}
\end{figure*}

\emph{Regime of strong driving}---Beyond the perturbative regime, the states $\{|A\rangle, |S_2\rangle, |A_2\rangle, |S\rangle \}$ are mixed by the driving and no longer represent the eigenstates of the system. In this situation, tractable analytical expressions for the eigenstates can only be obtained for limiting cases, e.g. $\beta=0$ or $\beta=\pi/2$. 

The bottom row of Fig.~\ref{fig:fig5-spectrum-omega} shows the eigenstates computed numerically for $\Omega = R$; the eigenstates are labeled $|U_i\rangle$, with $i=1\ldots 4$, in order of decreasing energy. The composition of the eigenstates depends on $\beta$, and for the particular cases $\beta = \{0,\pi/4,\pi/2\}$, it can be written as follows:

\begin{enumerate}
	\item $(\beta=0, \Delta=0)$. In this case, only the one-photon and two-photon symmetric states $|S\rangle$ and $|S_2\rangle$ hybridize due to the strong coherent drive. In this case, the eigenstates can be obtained analytically. To make the resulting expressions more readable, we write them here in non-normalized form:
	\begin{align}
		 |U^1 \rangle &\propto  \frac{R+\sqrt{R^2+16\Omega^2}}{4\Omega} |S \rangle + |S_2 \rangle , \\  
		| U^2 \rangle &= |A_2 \rangle ,\\
		| U^3 \rangle &= |A \rangle ,\\		
		| U^4 \rangle &\propto \frac{R-\sqrt{R^2+16\Omega^2}}{4\Omega} |S \rangle + |S_2 \rangle.
	\end{align}
One can see that, in the limit $\Omega \ll R$, these equations tend to the perturbative basis used above, for $\beta=0$. These eigenstates have the following eigenvalues:
	\begin{align}
		E_1 &= \frac{1}{2}\left( R  + \sqrt{R^2 + 16 \Omega^2} \right), \\
		E_2 &= 0 ,\\
		E_3 &= -R ,\\		
		E_4 &= \frac{1}{2}\left( R  - \sqrt{R^2 + 16 \Omega^2} \right).
	\end{align}
	which, in principle, would yield six sidebands. As discussed in the perturbative analysis, destructive interference between the emission from both emitters make all the transitions involving the antisymmetric state $|A\rangle$ invisible. These expressions for the eigenvalues also show that the eigenstates $E_3$ and $E_4$ cross in energy when $\Omega = R/\sqrt 2$.

	\item $(\beta=\pi/4, \Delta=0)$. This limit is not easily tractable analytically, so we limit our discussion to a description of the results from numerical calculations. As $\beta$ increases, $|U_1\rangle$ remains mostly a superposition of the two symmetric states, with a very small component of $|A\rangle$, and $|U_3\rangle$ and $|U_4\rangle$ mix, meaning that $|A\rangle$ is no longer an eigenstate of the system. This is an important observation, since it explains why the perfect destructive interference occurring for $\beta=0$ for transitions involving $|A\rangle$ no longer take place, and 13 peaks are visible. The eigenstates have the following structure:
	\begin{align}
		| U^1 \rangle &\approx C_{1,S} |S \rangle +C_{1,S_2} |S_2 \rangle ,\\		
		| U^2 \rangle &= |A^2 \rangle, \\ 
		| U^3 \rangle &= C_{3,S} |S \rangle +C_{3,A} |A \rangle +C_{3,S_2} |S_2 \rangle,\\
		| U^4 \rangle &= C_{4,S} |S \rangle +C_{4,A} |A \rangle +C_{4,S_2} |S_2 \rangle .
	\end{align}
where the $C_{i,j}$ ($i=1,3,4$, $j=S,A,S_2)$ represent generic amplitudes whose numerically computed values can be seen in Fig.~\ref{fig:fig6-spectrum-Delta}(g).
	
\item $(\beta=\pi/2, \Delta=0)$ This is the limit of detuned, uncoupled emitters, that recovers the physics of two independent, detuned Mollow triplets. The eigenstates have the following analytical form:
\begin{align}
	| U^1 \rangle &\propto C_+ |S \rangle + \frac{1}{2 C_+}|A\rangle +|S_2 \rangle ,\\		
	| U^2 \rangle &= |A_2\rangle , \\  
	| U^3 \rangle &\propto -\frac{\sqrt{2}\Omega}{R}|S \rangle +\frac{\sqrt{2}\Omega}{R} |A \rangle + |S_2 \rangle,\\
	| U^4 \rangle &\propto C_-|S \rangle +\frac{1}{2 C_-}|A \rangle + |S^2 \rangle .
\end{align}
with
\begin{equation}
C_\pm \equiv \frac{R\pm\sqrt{R^2+4\Omega^2}}{2\sqrt{2}\Omega}
\end{equation}
and eigenvalues
\begin{align}
E_1 &= \sqrt{R^2 + 4\Omega^2},\\
E_2 &= 0,\\
E_3 &= 0,\\
E_4 &= -\sqrt{R^2+4\Omega^2}.
\end{align}

\end{enumerate}

A clear conclusion of this analysis is that the spectrum resonance fluorescence is strongly sensitive to the ratio $J/\delta$ (expressed here in terms of the mixing angle $\beta$), and therefore it can serve as a valuable source of information about the parameters that characterize the system of quantum emitters, such as the inter-emitter distance. The continuous variation of the spectral features with $\beta$ is depicted in Fig.~\ref{fig:fig6-spectrum-Delta} for different values of $\Delta$. These plots reproduce the emergence and disappearance of peaks with $\beta$ described above, and furthermore show that similarly complex structures emerge outside the two-photon resonance, i.e. for $\Delta \neq 0 $, giving distinct patterns on the excitation-emission spectra that also depend strongly on $\beta$.

\section{Quantum parameter estimation of inter-emitter distances}
\label{sec:4-parameter-estimation}

\begin{figure*}[t]
	\begin{center}
		\includegraphics[width=0.98\textwidth]{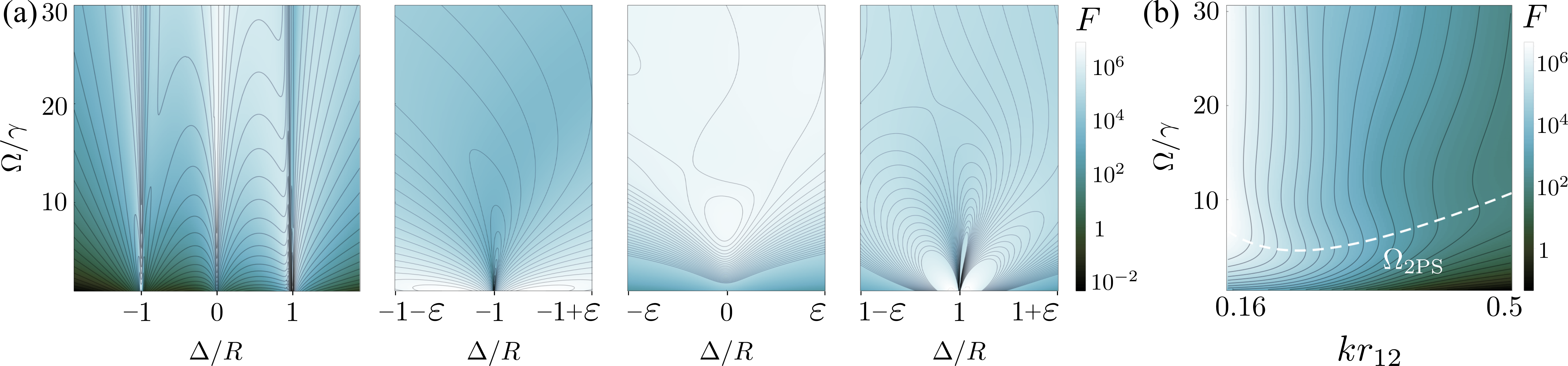}
	\end{center}
	\caption{Fisher information of resonance fluorescence measurements for the estimation of $k r_{12}$. (a) Fisher information in terms of $\Omega/\gamma$ and $\Delta$, for $k r_{12}=0.17$. The three rightmost panels depict a zoom around $\Delta = \{-R, 0, R \}$, with a zoom range is $\varepsilon = 0.02 R$. We observe that the region of two-photon excitation $\Delta\approx 0 $ features the higher Fisher information. (b) Fisher information versus $\Omega$ and $k r_{12}$ at $\Delta=0$. The white-dashed line represents the value of two-photon saturation amplitude $\Omega_{2\mathrm{PS}}$, which serves as a good indicator of the optimum driving amplitude that maximizes the Fisher information. Parameters: $\delta=50 \gamma$, $\Gamma = \gamma$, $\gamma_{12}=0.999\gamma$.  }
	\label{fig:fig7-fisher}
\end{figure*}

The results just discussed suggest that the measurement of excitation-emission spectra could provide valuable information for the estimation of internal parameters such as the value of the coherent coupling between emitters or, equivalently, their distance in real space. In this  section, we address this question by establishing the metrological potential of these measurements within the formalism of quantum parameter estimation~\cite{Luis2012,Chao2016,Dominik2018,Dowling2015,Liu2020,PARIS2009,PETZ2011,Wiseman2009}. To do this, we consider the precision achievable in the estimation of an unknown parameter, $X$, from the outcome of a positive operator-valued mesaurement (POVM) $\Lambda$, consisting of a set of operators $\{\hat\Lambda_\mu\}$, where the index $\mu\in\{1,2,\ldots,M\}$ denotes different possible measurement outcomes. The elements of the POVM add up to the identity, $\sum_\mu \Lambda_\mu = \mathbb{1}$. The probability for each of the different measurement outcomes follows a distribution $P(\mu|X) = \mathrm{Tr}[\rho_X \Lambda_\mu]$, where $\rho_X$ is the $X$-dependent density matrix of the system at the time of the measurement. When one infers $X$ from $P(\mu|X)$, the best attainable sensitivity $\Delta^2 X$ that can be achieved is given by the classical Fisher information of the probability distribution,
\begin{equation}
F=\mathrm E\left[\left(\frac{d\log P(\mu|X)}{dX}\right)^2\right].
\end{equation}
The Cram\'er-Rao bound then establishes that the minimum possible variance $\Delta^2 X$ in the estimation of $X$ is given by the inverse of $F$~\cite{Dowling2015},
\begin{equation}
\Delta^2 X \geq  1/F.
\end{equation}

Here, we will focus on the particular  example of the estimation of the inter-emitter distance $k r_{12}$ by measuring the fluorescence spectrum described in the previous section. In order to do this, we will make a series of assumptions. First, we will assume that the value of the spectrum at a frequency $\omega$ is obtained by performing a photon-counting measurement in a bosonic sensor resonant at that frequency and weakly coupled to the quantum emitters. For instance, this sensor can be understood as a tunable Fabry-Perot cavity used as a frequency filter. We assume that a discrete set of $N$ frequencies $\omega_n$ are measured. The corresponding POVM is given by the tensor product of photon-number operators of the different sensors, $\hat n_1\otimes \hat n_2 \ldots \otimes \hat n_N$, and the probability distribution that describes the measurement outcomes is of the form $P(n_1,n_2,\ldots,n_N|k r_{12})$, representing the joint photon-counting probability in each of the sensors. We then assume that the measurements done at different sensors are uncorrelated, so that $P(n_1,n_2,\ldots,n_N|k r_{12})=P(n_1|k r_{12})P(n_2|k r_{12})\ldots P(n_N|k r_{12})$. This ignores the possible contribution to the Fisher information of the correlations between photons emitted at different frequencies~\cite{DelValle2012,Gonzalez-Tudela2013a,SanchezMunoz2014a,LopezCarreno2017}. This assumption is justified if the lifetime of photons within the sensor is very long, so that temporal correlations are lost, or if different frequencies are measured sequentially in independent experiments, e.g. tuning the frequency of a Fabry-Perot filter. Following the same approach that is used in the theory of quantum image processing~\cite{Delaubert2008}, we assume that the resulting state of the sensor is a coherent state, so that the corresponding measured photo-current displays Poissonian fluctuations. This means that the photon-counting distribution associated to the sensor of frequency $\omega$ is given by
\begin{equation}
	P(n_\omega|k r_{12})=\frac{\langle n_\omega \rangle e^{-\langle n_\omega \rangle}}{n_\omega!},
	\label{eq:p_n_coherent}
\end{equation}
where $\langle n_\omega \rangle = \eta S(\omega)$, with $\eta$ a global constant that depends on the particular details of the detection scheme (e.g. detection efficiency). We set $\eta=1$ for simplicity, since it only yields an overall factor. Given that the spectrum $S(\omega)$ is strongly dependent on the value of $\beta$, it will also vary strongly with the the inter-emitter distance $k r_{12}$, which we emphasize by writing $S(\omega) = S(\omega,k r_{12})$. Since different sensors are uncorrelated and probability distributions factorize, the Fisher information associated to the measurement of the spectrum $S(\omega,k r_{12})$  is given by the sum  
\begin{multline}
F = \sum_\omega \mathrm E\left[\left(\frac{\partial \log P(n_\omega |k r_{12})}{\partial k r_{12}}\right)^2\right] \\= \sum_\omega\frac{1}{S(\omega,kr_{12})}\left[\frac{\partial S(\omega,k r_{12})}{\partial k r_{12}} \right]^2,
\end{multline}
where we used Eq.~\eqref{eq:p_n_coherent}. 
This quantity allows us to evaluate the metrological potential of fluorescence spectrum measurements. For this calculation, we take into account a finite detector linewidth $\Gamma = \gamma$ in the spectrum,~\cite{DelValle2012}, done by the replacement $\omega\rightarrow \omega+i\Gamma$ in Eq.~\eqref{eq:spectrum-equation}.
Our results are summarized in Fig.~\ref{fig:fig7-fisher}. Fig.~\ref{fig:fig7-fisher}(a) depicts the Fisher information versus the optically tunable parameters $\Delta$ and $\Omega$, showing that the optimal regime of operation is at the two-photon resonance, $\Delta\approx 0$, where $F$ is found to be larger. This is explained by the fact that the mechanism of two-photon excitation is strongly dependent on the coherent coupling between emitters, as we have seen in previous sections, and this is strongly modified by the inter-emitter distance $k r_{12}$. At $\Delta=0$, we find that there is an optimum value of the driving amplitude $\Omega$ that maximizes $F$ and consequently the precision in the estimation of $k r_{12}$. As we show in Fig.~\ref{fig:fig7-fisher}(b), this maximum varies with the actual value of $k r_{12}$, and it is well approximated by the driving amplitude of two-photon saturation $\Omega_{2\mathrm{PS}}$, that we obtained in Eq.~\eqref{eq:Omega_2PS}, at which the two-photon sidebands begin to be resolved in the spectrum. This establishes the onset of the two-photon saturation regime as the most sensitive point of operation for the estimation of the distance between interacting quantum emitters. We note that the results shown here as a function of the optically tunable parameters $\Delta$ and $\Omega$ represent different, independent experiments. Thus, a series of measurements over the $(\Omega,\Delta)$ parameter space could provide a higher precision of estimation, with a Fisher information that would be given by the sum $F=\sum_{\Delta,\Omega}F(\Omega,\Delta)$.

\section{Conclusions}
In this work, we have studied a system of two interacting, non-identical quantum emitters under coherent driving, focusing particularly on the regime of two-photon excitation. We have provided, for the first time for non-identical emitters, analytic approximations of the stationary density matrix and of steady-state observables such as the intensity of fluorescent emission. These calculations provide valuable insights on how the properties of the light emitted depend on degree of the coherent coupling and the detuning between emitters, and allows us to establish the regime of parameters in which specific features of two-photon excitation, such as the characteristic two-photon resonance peak in the excitation spectrum or two-photon sidebands in resonance fluorescence, are visible. Given that this features are strongly dependent on the coupling strength between emitters, and thus on their relative distance, we have explored the potential of these optical measurements for the estimation of the inter-emitter distance in terms of their Fisher information. We have established that the onset of two-photon effects at the two-photon resonance is the most sensitive point of operation for the estimation of the inter-emitter distance, a result that can be of great relevance for the problem of  imaging beyond Abe's resolution limit~\cite{Hettich2002}. 

\acknowledgements
The authors are thankful to D. Martin-Cano for insightful discussions. The project that gave rise to these results received the support of a fellowship from la Caixa Foundation (ID 100010434), from the European Union's Horizon 2020 Research and Innovation Programme under the Marie Sklodowska-Curie Grant Agreement No. 47648, with fellowship code  LCF/BQ/PI20/11760026.

\let\oldaddcontentsline\addcontentsline
\renewcommand{\addcontentsline}[3]{}
\bibliography{Collection,Books,library,sci-url}
\end{document}